\documentclass[twocolumn,superscriptaddress,amsmath,amssymb,longbibliography]{revtex4-1}

\usepackage{graphicx}
\usepackage{dcolumn}
\usepackage{bm}
\usepackage{subfigure}
\usepackage{amsmath}
\usepackage{appendix}
\usepackage{enumitem}

\newcommand{\qo}[1]{``#1''}                               
\newcommand{\op}[1]{\widehat{#1}}                   
\newcommand{\ket}[1]{|#1\rangle}                      
\newcommand{\bra}[1]{\langle #1|}                     
\newcommand{\me}[1]{\langle #1\rangle}    	
\newcommand{\brck}[1]{\left[ #1 \right]}		
\newcommand{\brc}[1]{\left\{ #1 \right\}}		
\begin{document}
\title{Dynamical moments reveal a topological quantum transition in a photonic quantum walk}
\author{Filippo Cardano}
\affiliation{Dipartimento di Fisica, Universit\`{a} di Napoli Federico II, Complesso Universitario di Monte Sant'Angelo, Napoli, Italy}
\author{Maria Maffei}
\affiliation{Dipartimento di Fisica, Universit\`{a} di Napoli Federico II, Complesso Universitario di Monte Sant'Angelo, Napoli, Italy}
\author{Francesco Massa}
\altaffiliation[Current address: ]{Faculty of Physics, University of Vienna, Boltzmanngasse 5, 1090, Vienna, Austria}
\affiliation{Dipartimento di Fisica, Universit\`{a} di Napoli Federico II, Complesso Universitario di Monte Sant'Angelo, Napoli, Italy}
\author{Bruno Piccirillo}
\affiliation{Dipartimento di Fisica, Universit\`{a} di Napoli Federico II, Complesso Universitario di Monte Sant'Angelo, Napoli, Italy}
\affiliation{CNR-SPIN, Complesso Universitario di Monte Sant'Angelo, Napoli, Italy}
\author{Corrado de Lisio}
\affiliation{Dipartimento di Fisica, Universit\`{a} di Napoli Federico II, Complesso Universitario di Monte Sant'Angelo, Napoli, Italy}
\affiliation{CNR-SPIN, Complesso Universitario di Monte Sant'Angelo, Napoli, Italy}
\author{Giulio De Filippis}
\affiliation{Dipartimento di Fisica, Universit\`{a} di Napoli Federico II, Complesso Universitario di Monte Sant'Angelo, Napoli, Italy}
\affiliation{CNR-SPIN, Complesso Universitario di Monte Sant'Angelo, Napoli, Italy}
\author{Vittorio Cataudella}
\affiliation{Dipartimento di Fisica, Universit\`{a} di Napoli Federico II, Complesso Universitario di Monte Sant'Angelo, Napoli, Italy}
\affiliation{CNR-SPIN, Complesso Universitario di Monte Sant'Angelo, Napoli, Italy}
\author{Enrico Santamato}
\affiliation{Dipartimento di Fisica, Universit\`{a} di Napoli Federico II, Complesso Universitario di Monte Sant'Angelo, Napoli, Italy}
\author{Lorenzo Marrucci}
\affiliation{Dipartimento di Fisica, Universit\`{a} di Napoli Federico II, Complesso Universitario di Monte Sant'Angelo, Napoli, Italy}
\affiliation{CNR-SPIN, Complesso Universitario di Monte Sant'Angelo, Napoli, Italy}
\maketitle
\label{sec:main}
\textbf {
Many phenomena in solid-state physics can be understood in terms of their topological properties \cite{Xiao10_RevModPhys,Qi11_RevModPhys}. Recently, controlled protocols of quantum walks are proving to be effective simulators of such phenomena \cite{Rudn09_PRL,Kita10_PRA,Kita12_NatCom,Asbo12_PRB}. Here we report the realization of a photonic quantum walk showing both the trivial and the non-trivial topologies associated with chiral symmetry in one-dimensional periodic systems, as in the Su-Schrieffer-Heeger model of polyacetylene \cite{SSH_PRL}. We find that the probability distribution moments of the walker position after many steps behave differently in the two topological phases and can be used as direct indicators of the quantum transition: while varying a control parameter, these moments exhibit a slope discontinuity at the transition point, and remain constant in the non-trivial phase. Extending this approach to higher dimensions, different topological classes, and other typologies of quantum phases may offer new general instruments for investigating quantum transitions in such complex systems. 
} \\

\begin{figure*}[H!]
\begin{center} 
\includegraphics[width=17cm]{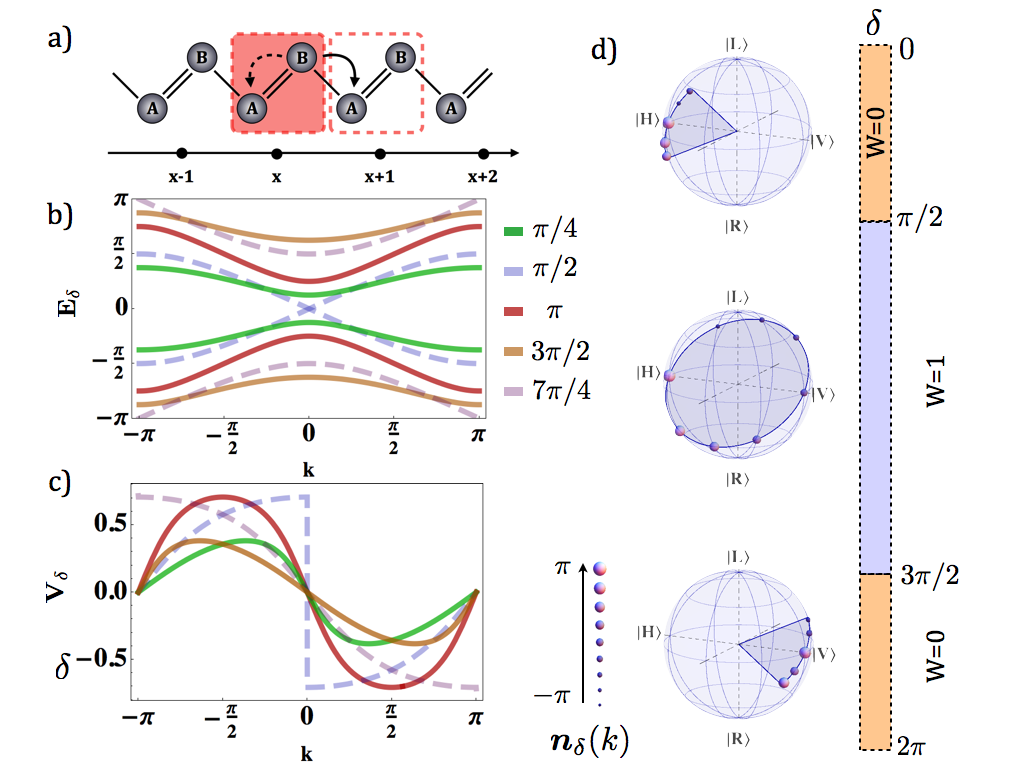} 
\caption{Topological characterization of the QW process and SSH model of the polyacetylene chain. a) Scheme of the polyacetylene chain in the SSH model. The latter is described as a sequence of two species of carbon atoms, that we denote as $A$ and $B$, which are linked by alternating single and double bonds. The unit cell is made of a pair of $A$ and $B$ atoms, whose repetition gives rise to the discrete lattice for the electron dynamics. The couplings between $A$ and $B$ sites in the same cell or in neighbor ones are represented by dashed and continuos arrows, respectively. The ratio of the corresponding transition amplitudes determines the topological features of the system. b-c) Dispersion relation for the quasi-energy $E_\delta (k)$ and the group velocity $V_\delta(k)$ of the Bloch bands in the QW. The dispersion curves for the two bands depend on the external parameter $\delta$; being the temporal coordinate a discrete variable, the quasi-energy is defined in a Brillouin zone $\{-\pi,\pi\}$. Here we report few examples; colors of the dispersion curves are associated with specific values of $\delta$, as shown in the panel legend. Generally, a finite gap separates the energies of the two bands; only at the two points $\delta=\delta_1=\pi/2$ and $\delta=\delta_2=3\pi/2$ (blue and purple dashed lines, respectively) the gap vanishes for $k=0$ and $k=\pi$, respectively, causing the presence of a discontinuity in the group velocity [see panel c)]. As shown in panel d), points $\delta_1$ and $\delta_2$ represent the boundaries between two topological phases. d) Topology of the Bloch eigenstates. States $\ket{\phi_s(k)}$, representing the polarization (coin) part of the eigenstates of the system, can be represented as points on the surface of a Poincar\'e sphere, where they are individuated by $\pm\boldsymbol{n}_\delta(k)$. Here we represent these states as solid spheres, whose radius is related to the value of $k$, as shown in the underlying legend. As the quasi-momentum $k$ varies in the Brillouin zone $\{-\pi,\pi\}$, chiral symmetry forces these states to lie on a great circle of the sphere. The associated winding number $W$ depends on $\delta$, and determines the existence of a non-trivial and a trivial topological phases; the former ($W=1$) occurs when $\delta_1<\delta<\delta_2$, while the latter ($W=0$) is obtained in the remaining part of the interval $\{0,2\pi\}$. As an example, from the top to the bottom of the panel, we represent the coin eigenstates of a QW with $\delta=\pi/4,\,\pi,\,7\pi/4$, respectively. For $\delta=\pi$ these form a closed loop along the great circle, while in the other cases they go back and forth along a finite arch, whose lengths depends on the value of $\delta$ (when $\delta=\{0,2\pi\}$, the arch results in a single point).}
\label{fig:QW}
\end{center}
\end{figure*}

The presence of topological order in matter is responsible for fundamental phenomena, such as the fractional and integer quantum Hall effects \cite{Thou82_PRL,Zhan05_Nat} or the existence of protected surface states in topological insulators \cite{Hasa10_RevModPhys}. Non-trivial topological phases are related to symmetries and can be characterized by specific topological invariants, defined in terms of the energy band eigenstates. The properties of these topological phases can be conveniently studied using quantum simulators \cite{Kita12_NatCom,Gome12_Nat,Atal13_NatPhys,Gens13_PRL,Hauk12_PRL,Zeun14_Axv}. In this context, quantum walks (QW) are emerging as a versatile tool \cite{Kita10_PRA}, since this simple quantum dynamics can realize all topological phases occurring in one- and two-dimensional systems of non interacting particles; as a first application of this concept, the formation of topologically protected bound states was observed in a photonic QW \cite{Kita12_NatCom}. 
In this paper we report the study of a novel QW process that exhibits both the trivial and non-trivial topological phases associated with chiral symmetry in a unidimensional (1D) bipartite lattice. A remarkable example of a physical system having identical symmetries and topological features is given by the polyacetylene chain, whose mobile electron dynamics is well described by the Su-Schrieffer-Heeger model (SSH) \cite{SSH_PRL}. SSH involves electron hopping between adjacent sites $A$ and $B$, which may belong to the same cell or not, as shown in Fig.\ \ref{fig:QW}a. Varying the relative strength of these two inter-site couplings, one predicts the existence of two distinct phases, marked by different values of a suitable topological invariant $W$. This topological classification is possible thanks to the presence of the chiral (or sublattice) symmetry associated with the interchange of the sublattices $A$ and $B$ \cite{Kita10_PRA,Asbo12_PRB,Asbo13_PRB} (see Supplementary Information [SI] for further details).
By studying our QW system, we found that the probability distribution moments for the particle position after a long temporal evolution show a different asymptotic behavior in the two topological phases: varying an external control parameter, these moments exhibit a slope discontinuity at the quantum transition, and remain locked to a constant value in the non-trivial topological phase. These effects occur also in the SSH model and hence seem to be fairly general, at least in the class of chiral-symmetric systems. We simulated experimentally such phenomena and propose a theoretical interpretation in terms of the dispersion relations and of the geometric features of the system eigenstates. Remarkably, our analysis takes into account bulk dynamics only, while not considering physical effects manifesting at the edges; the latter are the typical signature of topological systems. In the same spirit, topological phases have been recently detected in a non-Hermitian quantum walk \cite{Rudn09_PRL,Zeun14_Axv}.\\

A 1D quantum walk is the discrete evolution of a particle (the walker) on a 1D lattice \cite{Vene12_QIP}. At each step, the walker can move to the two nearest-neighbour sites; the direction of the shift is determined by the configuration of an internal 2-state quantum system (the coin). Between consecutive steps, the coin state undergoes a rotation, mimicking the coin toss characterizing the familiar classical random walk. 
QWs have been implemented in a variety of physical architectures \cite{Wang13}. In our photonic platform \cite{Card15_SciAdv}, the walker is encoded in the orbital angular momentum (OAM) of light \cite{Yao11_AOP}: discrete positions on the lattice are associated with states $\ket{m} $, where $m$ is an integer, describing a photon carrying $m\hbar$ of OAM along its propagation axis. The coin is encoded in the polarization degree of freedom: vectors $\ket{L}$ and $\ket{R}$, representing left and right circular polarizations, respectively, are the two internal states that determine opposite shift directions in the lattice.  The quantum state of the photon after $n$ steps of QW is denoted as $\ket{\psi_n}=\hat U_0^n\ket{\psi_0}$, where $\ket{\psi_0}$ is the initial (input) state and 
$\hat U_0$ is the unitary evolution operator of a single QW step (its formal expression is in Eq.\ \ref{ev_operator}). This evolution operator is realized with two optical elements: a quarter-wave plate (QWP) and a $q$-plate (QP) (Eqs.\ \ref{ev_operator}-\ref{qdelta1}). A $q$-plate is essentially a liquid-crystal birefringent cell having the optic axis arranged in a singular pattern \cite{Marr06_PRL}, with topological charge $q$ (in our case $q=1/2$). This patterned birefringence gives rise to an engineered optical spin-orbit coupling that induces the polarization-controlled shift of OAM. As specified in Eq.\ \ref{qdelta1}, besides the charge $q$, the action of this device is determined by the value of the optical retardation $\delta$, which can be tuned with an applied electric field \cite{Picc10_APL}.
A topological characterization of QWs with chiral symmetry may be defined in terms of the eigenstates of the operator $\hat U_0$ (see the SI) which, due to the lattice translation symmetry, are associated with two quasi-energy bands parametrized by a quasi-momentum $k$, as shown in Fig.\ \ref{fig:QW}b (see also Eq.\ \ref{dispersionQW}). As for momentum $k$, a periodic quasi-energy variable $E$ replaces here the usual energy as a consequence of time (= step number) being discretized. The coin part of the band eigenstates is denoted as $\ket{\phi_{s,\delta}(k)}$, where $s\in\{1,2\}$ is the band index. Conveniently, these states can be represented as points on the Poincar\'e sphere for light polarization, individuated by a three-dimensional unit vector that we refer to as $\boldsymbol{n}_\delta(k)$ (see Eq.\ \ref{eq:PTQW_eigenvectors}).
Chiral symmetry constrains vectors $\boldsymbol{n}_\delta(k)$ to lie on a great circle of the sphere. When $k$ varies throughout the Brillouin zone $\{-\pi,\pi\}$, the number of closed loops (winding number) described by $\boldsymbol{n}_\delta(k)$ is the topological invariant $W$. As shown in Fig.\ \ref{fig:QW}d, the value of $\delta$ in the range $\{0,2\pi\}$ determines the existence of two different topological phases: a non-trivial phase with $W=2q=1$ occurs when $\delta_1<\delta<\delta_2$, while $W=0$ (trivial phase) in the remaining region. In Fig.\ \ref{fig:QW}b it can be noted that when $\delta = \delta_1$ and $\delta = \delta_2$ the two bands are touching and the dispersion is locally linear at $k=0$ and $k=\pi$, respectively, as typically occurs for quantum transitions between topological phases \cite{Qi11_RevModPhys}. Interestingly, comparing the expressions of group velocity $V_\delta=\text d E_\delta/\text d k$ (\ref{eq:vgroup}) and vector $\boldsymbol{n}_\delta(k)$ (\ref{eq:PTQW_eigenvectors}), we can observe that $n_z=-n_y=V_\delta$. Thus, in the non-trivial phase, the maximum of the group velocity is independent of $\delta$ (see Fig.\ \ref{fig:QWdisp_detail} in the SI for further details). A recently introduced QW protocol, that is the Split-Step QW, has very similar properties \cite{Kita10_PRA,Kita12_NatCom}.\\

\begin{figure*}[H!]
\begin{center}
\includegraphics[width=17cm]{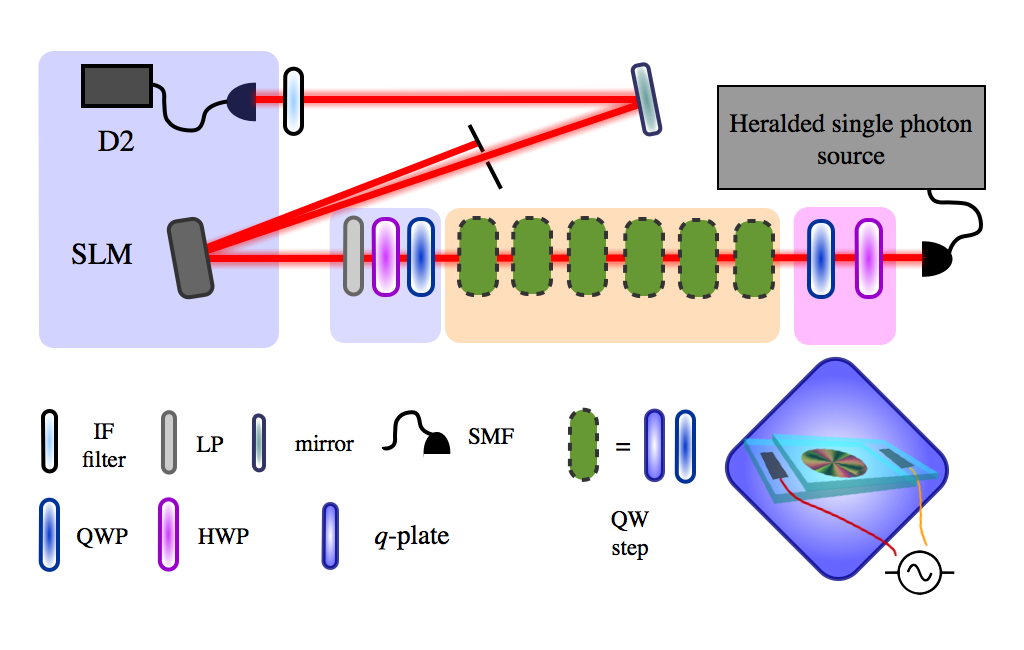} 
\caption{Layout of the setup. In the heralded single photon source (the optical components that are used to realize this source are not shown in the figure, for details see Ref.\ \cite{Card15_SciAdv}) laser pulses (100 fs) at 800 nm generated by a Titanium-Sapphire source (Ti:Sa) with repetition rate 82 MHz shine a type I Beta-Barium Borate crystal (BBO1) for second harmonic generation (SHG); frequency doubled pulses at 400 nm, with 110 mW average power and linear-horizontal polarization, pump a type II BBO crystal (BBO2), cut for collinear and degenerate spontaneous parametric down conversion (SPDC). Signal and idler photons, generated in horizontal and vertical linear polarizations, respectively, are spatially separated by means of a polarizing beam splitter (PBS) and then coupled to single-mode optical fibers (SMFs); the idler photon is directly sent to an avalanche photodiode (APD D1, not shown) while the signal one, after passing trough the QW system, is analyzed in polarization and OAM and finally detected by APD D2, in coincidence with D1. Before the QW, the photon exits the fiber in the OAM state $m=0$, and then its polarization is prepared in the state $\ket{\phi_0}=\alpha\ket{L}+\beta\ket{R}$; the two complex coefficients $\alpha$ and $\beta$ (with $|\alpha|^2+|\beta|^2=1$) are selected by using a half-wave plate (HWP) and a quarter-wave plate (QWP) (apart from an unimportant global phase). After the initial state preparation, the photon goes through the 6-step QW, with the single step consisting of a $q$-plate and a QWP oriented at 90$^\circ$. For each $q$-plate, the value of the optical retardation $\delta$ is controlled by the amplitude of an alternating electric field, which is introduced by means of an external generator \cite{Picc10_APL}. At the exit of the QW, a polarization projection is realized using a second HWP-QWP set followed by a linear polarizer (LP). The OAM state is then analyzed by diffraction on a spatial light modulator (SLM), followed by coupling into a SMF. Before detection, interferential filters (IF) centered at 800 nm and with a bandwidth of 3.6 nm are used for spectral cleaning. The latter was required since the photons wavelength strongly influence the action of the devices implementing the QW ($q$-plate and QWP).}
\label{fig:setup}
\end{center}
\end{figure*}

\begin{figure*}[H!]
\begin{center}
\includegraphics[width=17cm]{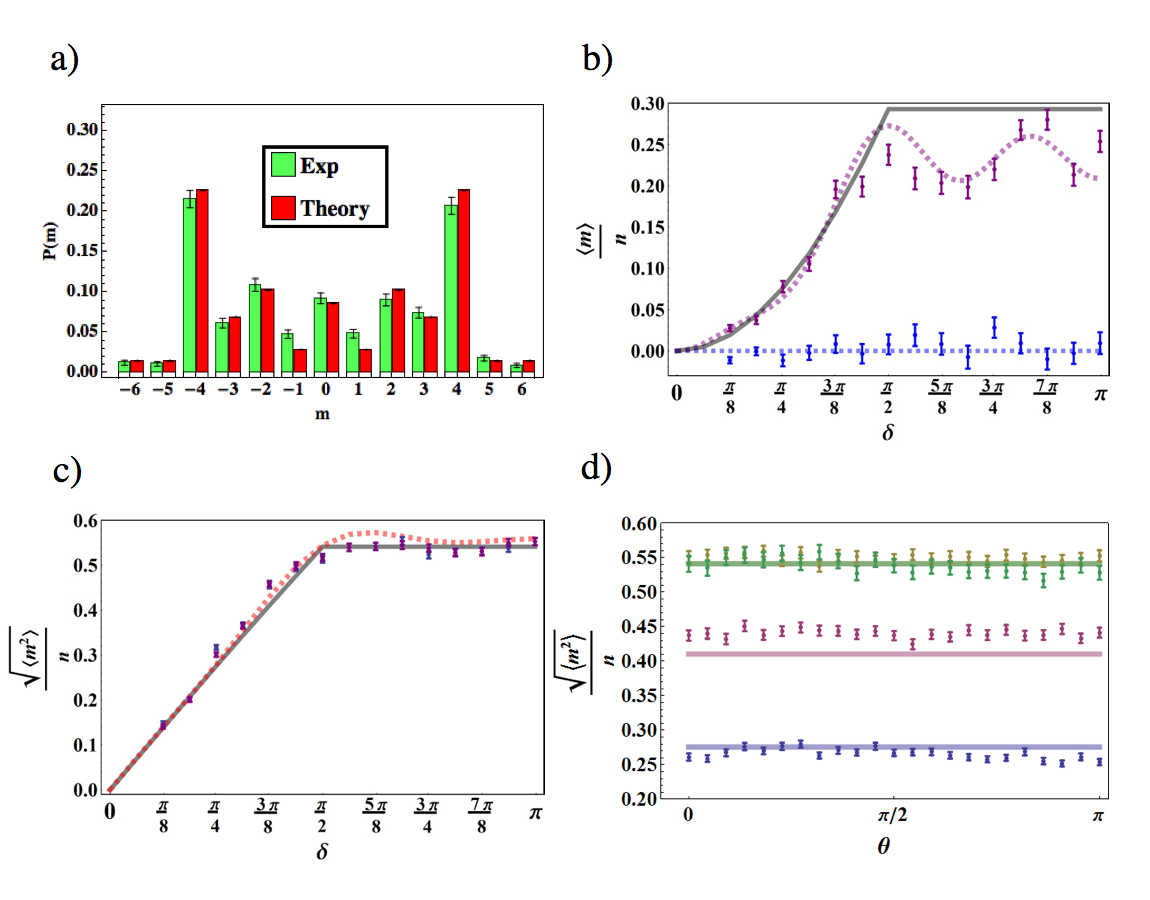} 
\caption{Experimental results. a) Example of a probability distribution for the walker. Measured (green, left) and expected (red, right) probability distributions after a 6 steps QW of a photon initially prepared in the state $m=0$ and $\brc{\alpha,\beta}=1/\sqrt 2\brc{1,1}$, when $\delta=2.95$. The error bars represent statistical errors at one standard deviation, calculated assuming Poissonian fluctuations on single counts. b-c) Measured values of statistical moments $M_1/n$ and $\sqrt{ M_2}/n$, respectively, when varying $\delta$ in the range $\{\pi/8,\pi\}$ with steps of $\pi/16$, for initial states $\brc{\alpha,\beta}=1/\sqrt 2\brc{1,1}$ (blue dots), and $\brc{\alpha,\beta}=\brc{0,1}$ (purple dots). Dashed lines represent expected values for the plotted quantities, as obtained from numerical simulations. The continuous lines are the asymptotic limits reported in Eq.\ \ref{eq:PTQW_M2}.  d) Measured values of $\sqrt{M_2}/n$ when the initial polarization corresponds to the state $\cos(\theta/2)\ket{L}+\sin(\theta/2)\ket{R}$, varying the polar angle $\theta\in[0,\pi]$ with steps of $\pi/22$; data are collected in correspondence of $\delta=\pi/4$ (blue), $\delta=3\pi/8$ (purple), $\delta=3\pi/4$ (yellow), $\delta=\pi$ (green). For each of these configurations, continuous lines give the corresponding asymptotic limit obtained using Eq.\ \ref{eq:PTQW_M2}. When $\delta=3\pi/8$, experimental data are slightly higher then the associated continuous line; nevertheless, the discrepancy  is compatible with the results of numerical simulations for a finite number of steps (see Fig.\ \ref{fig:PTQW_M12}d). These data show that $M_2$ is asymptotically independent of the input coin-polarization state.}
\label{fig:data}
\end{center}
\end{figure*}

The features of the energy bands and associated eigenstates have profound consequences on the system dynamics, which shows marked differences in the two topological phases \cite{Rudn09_PRL,Zeun14_Axv}. Here we characterize such different behavior through the analysis of the moments of the probability distribution $P(m)$ associated with the walker position, defined as $M_j=\sum_m\,m^j\,P(m)$. In particular, we consider a photon starting its walk in the position $m=0$ with an arbitrary polarization, that is $\ket{\psi_0} = \ket{0}\otimes\ket{\phi_0}$, where $\ket{\phi_0}$ is a generic state in the coin Hilbert space. Simulations of the QW evolution in the large step-number limit show that the moments $M_j$ assume a constant value (independent of $\delta$) in the non-trivial phase, and undergo abrupt variations at the phase transitions, that is at $\delta=\{\delta_1,\delta_2\}$. In the infinite-steps-limit, these moments have simple asymptotic expressions in terms of the energy band dispersion relations; in particular, those relative to the first and second moments $M_1$ and $M_2$ are the following (see the SI for a proof):
\begin{align}
M_{1} /n&=  (s_{y}-s_{z}) L(\delta)+ O(1/n)\label{eq:PTQW_M1}\\
M_{2}/n^2&= L(\delta)+ O(1/n^2)\label{eq:PTQW_M2}
\end{align}
where $s_i=\bra{\phi_0}\hat \sigma_i\ket{\phi_0}$, with $i\in\{x,y,z\}$, are the expectation values of the Pauli operators $\op \sigma_i$ in the coin space, calculated with respect to coin initial state $\ket{\phi_0}$. Interestingly, we observe that $M_2$ is independent of the initial coin state. The quantity $L(\delta)$ appearing in Eqs.\ \ref{eq:PTQW_M1},\ref{eq:PTQW_M2} is equal to the square of the group velocity $V_\delta$, and hence to the square of $n_y$, integrated in momentum space over the Brillouin zone:
\begin{align}\label{eq:PTQW_L}
L(\delta)=\int_{-\pi}^\pi \frac{\text d k}{2\pi}\left[V_{\delta}(k)\right]^2=\int_{-\pi}^\pi \frac{\text d k}{2\pi}\left[n_y(k,\delta)\right]^2.
\end{align}
For our QW process, the integral reported in Eq.\ \ref{eq:PTQW_L} admits a closed form (Eq.\ \ref{eq:Ldeltafun}). In Fig.\ \ref{fig:data}b, the quantity $L(\delta)$ is plotted as a function of $\delta$; as a consequence of the discontinuity present in the group velocity, $L(\delta)$ is a piecewise function, and manifests abrupt variations at $\delta_1$ and $\delta_2$. In the non-trivial phase, it is equal to a constant; this result can be qualitatively understood by looking at how the group velocity dispersion is modified by a change of $\delta$ (Fig.\ \ref{fig:QWdisp_detail}). On the other hand, in the trivial phase $L(\delta)$ increases or decreases in the two regions $\{0,\pi/2\}$ and $\{3\pi/2,2\pi\}$, respectively. A similar behavior occurs in the SSH model (see Fig.\ \ref{fig:SSH_res}) and hence seems to be a general feature of the non-trivial topology. In a QW with a finite number of steps, statistical moments $M_1$ and $M_2$ have a continuous behavior, converging to that given by Eqs.\ \ref{eq:PTQW_M1}-\ref{eq:PTQW_M2} asymptotically as $n\rightarrow \infty$. For $M_2$, this convergence is rapid and visible for values of $n$ that are small enough to be achieved in an experimental simulation, whereas for $M_1$ such process is much slower (see Fig.\ \ref{fig:PTQW_M12}). Thus, as a figure of merit for the quantum transition, we chose to analyze $\sqrt{M_2/n^2}=\frac{\sqrt{\me{\hat{ m}^2}}}{n}$. In particular, we implemented a 6-step QW and measured the latter quantity when varying the external parameter $\delta$, so as to simulate the topological quantum transition.\\
The layout of the apparatus is shown in Fig.\ \ref{fig:setup}. A standard \qo{heralded single photon source} is realized at the input of the QW system, as discussed for instance in Ref.\ \cite{Card15_SciAdv}. After the initial state preparation (see the caption of Fig.\ \ref{fig:setup} for more details) the photon undergoes the QW evolution and, at the exit, is analyzed in both polarization and OAM so as to determine the output probabilities and the associated moments. 
Letting $\delta$ vary in the range $\{0,\pi\}$ with steps of $\pi/16$, we determined the corresponding probability distribution of the walker after a 6-step QW; a typical example is reported in Fig.\ \ref{fig:data}a. 
In Fig.\ \ref{fig:data}c we plot the measured values for $\sqrt{M_2}/n$ as a function of $\delta$. In good agreement with quantum predictions, the data are close to the asymptotic limit reported in Eq.\ \ref{eq:PTQW_M2}; the emergence of an abrupt slope variation at $\delta=\pi/2$ can be appreciated and it is evident that in the non-trivial phase $M_2$ is constant. The observed discontinuity is a signature of the underlying quantum transition, while the existence of a constant value for $M_2$ in the non-trivial phase reflects the features of the group velocity dispersion, and of the geometry and topology of the system eigenstates.
Preparing the initial state of the coin ($m$=0 for the walker) in two non-orthogonal polarizations we also verified that $M_2$ (unlike $M_1$) is independent of the coin initial conditions, as expected from Eq.\ \ref{eq:PTQW_M2} [see Figs.\ \ref{fig:data}b-c)]. This aspect was further investigated by measuring $\sqrt{M_2}/n$ for several initial polarizations, corresponding to specific points along a meridian of the Poincar\'e sphere, and repeating the experiment for two values of $\delta$ in each topological sector. The final data, reported in Fig.\ \ref{fig:data}d, match well the predicted results and clearly show that the value of the second-order moment is independent of the coin initial state.\\

In conclusion, we have shown that a signature of quantum transitions between distinct topological phases is present in the behavior of suitable dynamical observables that are easily accessed experimentally. In the context of 1D systems with chiral symmetry, such as the SSH model, we propose that the statistical moments of the particle probability distribution in space in the large step-number limit provide a convenient choice of such observables. We experimentally validated this proposal by simulating this topological environment within a specific photonic QW architecture. The measured asymptotic moments as a function of a control parameter remain constant in the non-trivial topological phase and show slope discontinuities at the phase change. We also proved that these statistical quantities can be simply related to properties of the energy bands and in particular are directly linked to the square of the group velocity, integrated over the Brillouin zone. The latter, in turn, can be traced back to the geometric features of the system eigenstates (although not just to the associated topological invariant), thus providing insight on the general features leading to the emergence of such phenomena.
In prospect, our approach based on dynamical moments could be applicable to the investigation of other classes of quantum transitions and to topological phases associated with more complex symmetries or with a higher dimensionality, thus helping to shed new light on the physics of topological phenomena and on their simulation in suitable experimental architectures.
\bibliographystyle{unsrt}

\vspace{1 EM}
\noindent\textbf{Acknowledgments}\\
\noindent We thank Maciej Lewenstein, Pietro Massignan and Alessio Celi for preliminary reading of the manuscript and for providing useful comments, and Domenico Paparo and Antonio Ramaglia for lending some equipment. This work was partly supported by the Future Emerging Technologies FET-Open Program, within the 7$^{th}$ Framework Programme of the European Commission, under Grant No.\ 255914, PHORBITECH.
\vspace{1 EM}

\noindent\textbf{Author Contributions}\\ 
\noindent F.C., M.M, F.M., G.D.F., V.C., E.S. and L.M. devised various aspects of the project and designed the experimental methodology. F.C. and M.M., with contributions from C.d.L., carried out the experiment and analyzed the data. B.P. prepared the q-plates. F.C., M.M. and L.M. wrote the manuscript, with contributions from E.S., G.D.F and V.C.. All authors discussed the results and contributed to refining the manuscript.
\vspace{1 EM}

\noindent\textbf{Competing interests}\\
The authors declare no competing financial interests. Correspondence and requests for materials should be addressed to L.M. (lorenzo.marrucci@unina.it).
\clearpage

\onecolumngrid
\appendix
\noindent\textbf{SUPPLEMENTARY INFORMATION}
\vspace{1 EM}
\section{Role of the external parameter $\delta$ in tailoring the structure of the energy band of the QW system.}\label{SI:0}
\begin{figure}[h]
\begin{center} 
\includegraphics[width=13cm]{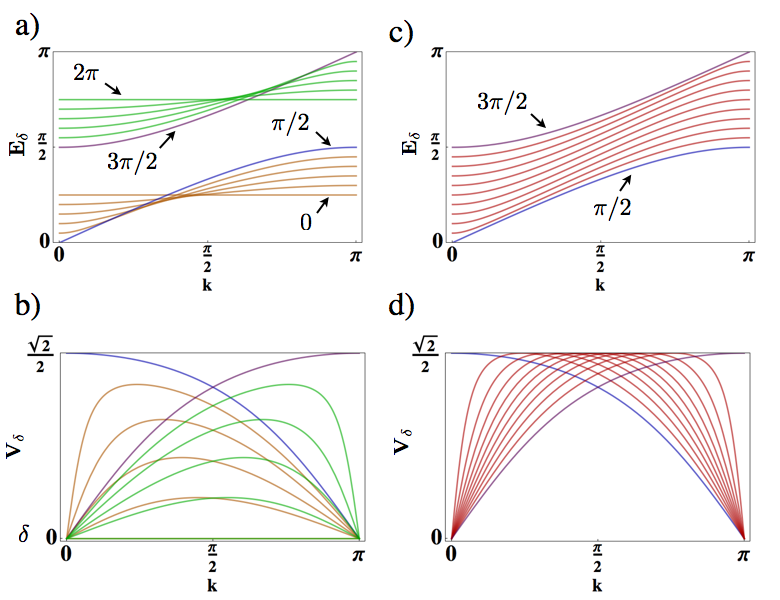} 
\caption{The parameter $\delta$ and the dispersion relations of our QW system. We plot the dispersion relation for the upper energy band, and the associated dispersion of the group velocity, of a QW system described by the single step operator $\op U_0$, with varying the external parameter $\delta$ in the range $\{0,2\pi\}$ with steps of $\pi/10$; in these plots, we restrict our attention to half of the Brillouin zone, being the dispersion relation symmetric in the other region ($E(k)=E(-k)$). In all images, coloured blue and purple curves correspond to the two configurations in which the quantum transition occurs, that is $\delta=\pi/2$ and $\delta=3\pi/2$, respectively. In panels a-b), coloured orange and green curves correspond to $\delta$ being in the intervals $\{0,\pi/2\}$ and $\{3\pi/2,2\pi\}$, respectively. As a consequence of the energy bands properties, in the former case, as $\delta$ increases, the maximum of the group velocity (panel b) becomes larger; in the latter configuration, this behavior is reversed. In panel c-d), coloured red curves correspond to $\delta \in \{ \pi/2,3 \pi/2\}$. In this case, the group velocity maximum remains locked to a constant value (panel d), while the corresponding quasi-momentum is drifting. In all panels, the correspondence between values of $\delta$ and the associated plots reflects the order in which these curves are displayed.} 
\label{fig:QWdisp_detail}
\end{center}
\end{figure}

In this section, we discuss the dependance of the dispersion relations of our QW system with respect to the value of the parameter $\delta$. The explicit expression of the QW quasi-energies is given by the following relation
\begin{align}\label{dispersionQW}
E_\delta(k)=\pm\text{cos}^{-1}\brck{\frac{\cos{ \left(\delta/2\right)}+\sin{\left(\delta/2\right)}\cos{\left(k\right)}}{\sqrt 2}},
\end{align}
We recall that $E_\delta (k)$ are the eigenvalues of the effective Hamiltonian $\hat H_\text{eff}$; accordingly, they represent a quasi-energy defined in a Brillouin zone $\{-\pi,\pi\}$, as a consequence of the temporal coordinate (the step-number) being a discrete variable. The expression of the related group velocity, that is $V_\delta=\text d E_\delta/\text d k$, is given by
\begin{equation}\label{eq:vgroup}
V_\delta(k) = \frac{dE_\delta(k)}{dk} = \pm\frac{\sin(\delta/2)\cos(k)}{\sqrt{2-[\cos(\delta/2)+\sin(\delta/2)\sin(k)]^2}}.
\end{equation}
In Fig.\ \ref{fig:QWdisp_detail}, we plot the quasi-energy and the group velocity as a function of the quasi-momentum $k$, for different values of $\delta$ in both the trivial and the non-trivial phases. Here we restrict our attention to the upper band (the energy of the other has only opposite sign), and to half of the Brillouin zone, having the dispersion relation the symmetry $E(k)=E(-k)$. At first glance, it is clear that the energy bands have marked differences with respect to the system being in the trivial or the non-trivial topological phases. In panel a), we can observe that as $\delta$ goes from $0$ to $\pi/2$ (or from $3\pi/2$ to $2\pi$), the band covers a larger (smaller) range of energy values; accordingly, in panel c), we can observe that the maximum values for the group velocity increases (decreases). In contrast with what we observed in the trivial phase, in panel b) it can be noted that in the non-trivial phase a change of $\delta$ results in a shift of the whole band, plus a tiny deformation of the curve. In panel d), we can observe that this results in a alteration of the group velocity dispersion that keeps constant the associated maximum value; this can be seen as a consequence of the geometric properties of the unit vector $\boldsymbol{n}_\delta(k)$ in the non-trivial phase, being valid the relation $V_\delta=n_z=-n_y$. The consequences of these features on the statistical moments associated with a QW evolution in the large step-number limit have been deeply discussed in the main text.\\

\section{SSH model and QW system: analogies and differences}\label{SI:1}
A typical example of a 1D periodic system with an internal 2-state degree of freedom, characterized by chiral symmetry, is the Su-Schrieffer-Heeger model (SSH) describing the electron dynamics in the polyacetylene chain. We discuss explicitly this model to introduce the symmetries that characterize our QW, highlighting the analogies and the differences between these different systems. The SSH model describes the polyacetylene chain as a one-dimensional (1D) dimerized lattice \cite{SSH_PRL}, where each unit cell consists 
of two sites $A$ and $B$, as shown in Fig.\ 1a. Complex amplitudes $t$ and $t'$ quantify the hopping between adjacent sites, belonging to the same unit cell or not, respectively (for the purposes of our analysis it is sufficient to consider $t$ and $t'$ as real parameters). The Hamiltonian describing the electron dynamics along this dimerized lattice is typically introduced for a multi-particle system; accordingly we express it using the second quantization formalism, (please note that we adopted the typical choice to present the QW as a single particle model; in this context the two representations are completely equivalent):
\begin{align}\label{eq:hamiltonian_ssh}
\hat{H}= \left\{\sum_{n=1}^{N} t\,\hat{c}^{\dagger} _{A,n}\hat{c}_{B,n}+ \sum_{n=1}^{N-1} t^\prime\, \hat{c}^{\dagger} _{A,n+1}\hat{c}_{B,n}\right\}+\text{h.c.}.
\end{align}
Here, $\hat{c}^{\dagger} _{A(B),n}$ and $\hat{c} _{A(B),n}$ are the creation and annihilation operators for an electron in the cell $n$, on the sublattice $A$ ($B$), and h.c. stands for \qo{hermitian conjugate}. We are considering a finite chain made of $N$ cells, which are spaced by a distance $a$. Using the Fourier transform for the creation and annihilation operators
\begin{align}
\hat{a}_k &=\frac{1}{\sqrt{N}}\sum_n e^{-i k n a }\hat{c} _{A,n} \label{eq:trasf1},\\
\hat{b}_k &=\frac{1}{\sqrt{N}}\sum_n e^{-i k n a }\hat{c} _{B,n} \label{eq:trasf2}
\end{align}
we can express the Hamiltonian operator in momentum space:
\begin{align}\label{eq:hamiltonian_ssh_k}
\hat{H}= \sum_{k} \hat{\psi} ^{\dagger}_{k}\{ [ t + t^{\prime} \cos{(a\,k)} ]\,\hat{\sigma}_x + t^{\prime} \sin{(a\,k)}\,\hat{\sigma}_{y}\} \hat{\psi}_k.
\end{align}
In the latter equation, $\hat{\psi}_k$ is a 2D vector operator defined as
\begin{align}\label{eq:fourier} 
\hat{\psi}_k =\binom{\hat{a}_k}{\hat{b}_k},
\end{align} 
and $\hat\sigma_x$ and $\hat\sigma_y$ are the 2D Pauli operators. For a better comparison with the QW system, it is convenient to operate a shift of the quasi-momentum in the Brillouin zone, that is $k\rightarrow k+\pi$. After this transformation, the eigenvalues of the SSH Hamiltonian (\ref{eq:hamiltonian_ssh_k}) are given by:
\begin{align}\label{eq:energy_ssh}
E_{t,t^\prime}(k)=\pm \sqrt{\left[t^2+{t^\prime}^2-2 t t^\prime \cos{(a\,k)}\right]}.
\end{align}
Henceforth, we will assume $a=1$, so as to switch to adimensional units (that characterize intrinsically the quantum walk system). Using Eq.\ \ref{eq:energy_ssh}, we can determine the group velocity dispersion, which has the following expression;
\begin{align}\label{eq:vgroup_ssh}
V_{t,t^\prime}(k)=\pm\frac{t t^\prime \sin{k}}{ \sqrt{\left[t^2+{t^\prime}^2-2 t t^\prime \cos{(a\,k)}\right]}}.
\end{align}
In momentum  representation, the Hamiltonian can be conveniently expressed as a $2\times2$ matrix: 
\begin{align}
H(k)= E_{t,t^\prime}(k)\,\boldsymbol{n}_{t,t^{\prime}}(k)\cdot \boldsymbol \sigma
\end{align}
where $\boldsymbol \sigma$ is the 3D vector whose components correspond to the three Pauli matrices. At each momentum $k$, the expression of the real 3D unit vector that determines the position of the coin eigenstates on the Poincar� sphere representing the spin Hilbert space is:
\begin{align}\label{eq:vec_ssh}
 \boldsymbol{n}_{t,t^{\prime}}(k)=\left\lbrace \frac{t- t^{\prime} \cos{k}}{ \sqrt{t^2+t^{\prime 2}-2 t t^{\prime} \cos{k}}},-\frac{t^{\prime} \sin{k}}{ \sqrt{t^2+ t^{\prime 2} -2 t t^{\prime}\cos{k}}} ,0 \right\rbrace 
\end{align}
In the latter expression, we can observe that the $z$ component of $\boldsymbol{n}_{t,t^{\prime}}(k)$ is vanishing; this results from the absence of $A - A$ and $B - B$ links in the Hamiltonian. Accordingly, the vector $\boldsymbol{n}_{t,t^{\prime}}(k)$ is confined on the equator of the Poincar\'e sphere that we use to represent the 2D sublattice Hilbert space. Moreover, we can note that the components of $\boldsymbol{n}_{t,t^{\prime}}(k)$ are related to the group velocity dispersion; indeed we have that
\begin{align}\label{eq:n_V_SSH}
n_y=\frac{V_{t,t'}}{t}.
\end{align}

\begin{figure}
\begin{center} 
\includegraphics[width=13cm]{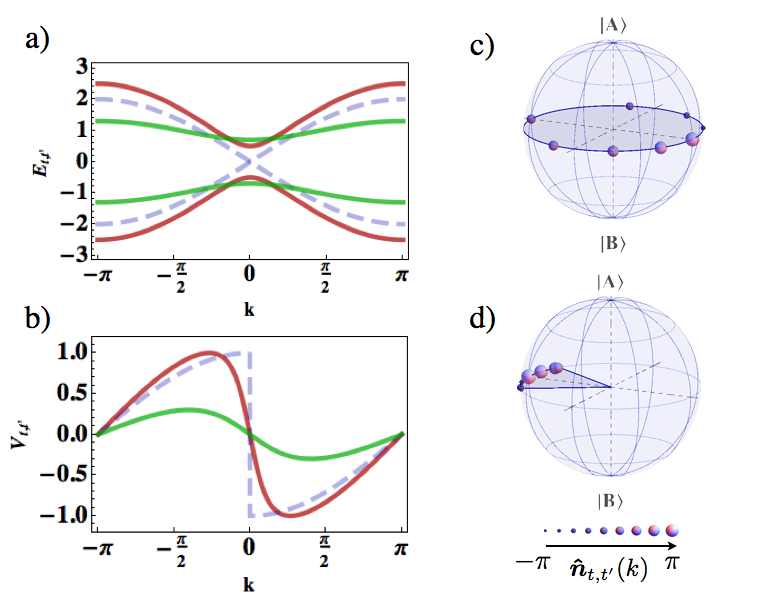} 
\caption{Dispersion relations and eigenstates for the SSH model. a-b) The values of the quasi-energy $E_{t,t'}$ for the two bands are reported as a function of the quasi-momentum $k$, in the Brillouin zone $\{-\pi,\pi\}$. Coloured red and green curves correspond to the two configurations $(t,t^\prime)=(1,1.5)$ and  $(t,t^\prime)=(1,0.3)$, respectively. The dashed curve corresponds to the case $t=t^\prime=1$. b) Group velocity dispersion for the lower energy band. Colour legend is the same with respect to panel a). From these plots, we can observe that the energy gap closure at $k=0$ (dashed curves) is associated with a jump of the group velocity. c-d) Representation of the vector $\boldsymbol{n}_{t,t^{\prime}}(k)$, calculated when varying $k$ in the range $\{-\pi,\pi\}$, with steps of $\pi/4$; hopping amplitudes $(t,t')$ are equal to $(2,1)$ and $(1,2)$, respectively. These vectors are represented by solid spheres, whose radius is associated with the quasi-momentum value, as reported in the legend. The coin eigenstates corresponding to the two configurations $t<t^\prime$ [panel c)] and $t>t^\prime$ [panel d)] have marked differences in terms of their topology. Indeed, when varying $k$ in the Brillouin zone, the associated trajectory over the Poincar\'e sphere results in closed loop or in a finite arch, respectively. Accordingly, the winding number is one in the first case, whereas it vanishes in the second one.}
\label{fig:SSH}
\end{center}
\end{figure}
In Fig.\ \ref{fig:SSH} we report the dispersion relations and the coin eigenstates for the SSH model, computed using Eqs.\ \ref{eq:energy_ssh} and \ref{eq:vec_ssh}, respectively, for different values of $t$ and $t^\prime$. In these plots we can see that the two bands have a finite gap, vanishing when $t=t^\prime$ (Fig.\ \ref{fig:SSH}a). Configurations $t<t^\prime$ and $t>t^\prime$ are not equivalent, being the winding number of $\boldsymbol{n}_{t,t^{\prime}}(k)$, as $k$ varies in the Brillouin zone $\{-\pi,\pi\}$, equal to one (Fig.\ \ref{fig:SSH}c) and zero (Fig.\ \ref{fig:SSH}d), respectively; this is the same topological invariant that we introduced for the QW. Similarly to the latter system, we can observe that the dispersion of the group velocity is different in the two distinct phases; the main features are the same presented in the previous section, in the context of a QW evolution.\\
Now, let us discuss briefly the symmetries characterizing the SSH system. First of all, the model has time-reversal symmetry: $\hat T\hat H\hat T^{-1}=\hat H$, where the time-reversal operator $\hat T$ is defined by $\hat T=\hat K$ and $\hat K$ is the complex conjugation operator. Indeed it is straightforward to show that $\hat H^{*}(k)=\hat H(-k)$. Moreover the system has the sublattice, or chiral, symmetry $\hat S\hat H(k)\hat S^{-1}=-\hat H(k)$. The sublattice symmetry operator $\hat S$ is defined by $\hat \sigma_z$, where $\hat \sigma_z$ is the z-component of the vector that represents the Pauli matrices and acts on the sub lattice degree of freedom. Finally the system has the particle-hole symmetry: $\hat P\hat H(k)\hat P=-\hat H(k)$, where the particle-hole operator is defined by $\hat P=\hat \sigma_z\hat K$. Then the topological class is BDI since 
$\hat P^2=\hat T^2=\hat S^2=1$ \cite{Ryu10_NJP,Schn08_PRB,Kita09_AIP,Altl97_PRB}. The class of 1D Hamiltonian operators with these symmetries is generally labeled as SSH, since the associated physical systems have the same topological features as those emerging in the SSH model. This is the case of the QW system we introduced, as we will demonstrate in the following paragraphs.\\
As explained in the main text, a QW consists in the dynamical evolution of a system made of a walker and a coin part; in our photonic platform, they are encoded in the spin angular momentum and in the orbital angular momentum of light, respectively. In particular, the unitary operator $\hat U_0$ defining one step of the quantum walk, consists of a quarter wave plate, oriented at $90^\circ$ with respect to the horizontal direction, followed by a $q$-plate. 
\begin{align}\label{ev_operator}
\hat U_0=\hat Q_\delta \hat W_\text{qwp}.
\end{align}
The QWP action is described by the operator $\hat W_\text{qwp}$, acting only on the polarization degree of freedom:
\begin{align}\label{op_qwp}
\hat W_\text{qwp}\ket{L,m} &=\frac{1}{\sqrt 2}\left(\ket{L,m}+i\,\ket{R,m}\right)\nonumber\\
\hat W_\text{qwp}\ket{R,m} &=\frac{-i}{\sqrt 2}\left(\ket{L,m}-i\,\ket{R,m}\right)
\end{align}
The $q$-plate action is described by the operator $Q_{\delta}$
 \begin{align}
\hat Q_{\delta}\ket{L,m} &=\cos{\left(\delta/2\right)}\ket{L,m}+i \sin{\left(\delta/2\right)}\ket{R,m+2q},\nonumber\\
Q_{\delta}\ket{R,m} &=\cos{\left(\delta/2\right)}\ket{R,m}+i \sin{\left(\delta/2\right)}\ket{L,m-2q}\label{qdelta1}.
\end{align}
Here, $q$ is the topological charge of the $q$-plate; $\ket{L}$ and $\ket{R}$ represent left 
circular and right circular polarizations; and $\delta$ denotes the optical birefringent 
phase-retardation. In our experiment, we considered $q=1/2$. It is possible to construct a time independent effective Hamiltonian, $\hat H_\text{eff}$, such that the quantum walk represents a stroboscopic simulator of the evolution generated by $\op H_\text{eff}$ at discrete times. By using  momentum eigenstates $\ket{k}$, $\op H_\text{eff}$ is given by:
\begin{align}\label{hamiltonian}
\op H_\text{eff}(\delta)=\int^{\pi}_{-\pi}dk\, \left[ E_{\delta}(k) \boldsymbol{n}_{\delta}(k) \cdot 
\boldsymbol{\sigma}\right]\otimes \ket{k} \bra{k},
\end{align}
where $E_{\delta}(k)$ represents the modulus of the Hamiltonian eigenvalues (see Eq.\ \ref{dispersionQW}), $\boldsymbol{\sigma}$ 
denotes the vector of Pauli matrices, and $\boldsymbol{n}_{\delta}(k)$ is a real 
3D unit vector representing the coin eigenstates on the Poincar\'e sphere associated with the spin Hilbert space. The components of this vector are given by
\begin{align}\label{eq:PTQW_eigenvectors}
n_x(k)&=\left\{\cos\left(\delta/2\right)- \sin\left(\delta/2\right)\cos k\right\}/N(k), \nonumber\\
n_y(k)&=- \sin\left(\delta/2\right)\sin k/N(k)\nonumber,\\
n_z(k)&=-n_y(k),
\end{align}
where the quantity $N$ is a normalization factor
\begin{align}\label{eq:PTQW_eigenvectors_N}
N(k)=\sqrt {2\left\{1-\cos^2[E_\delta(k)]\right\}}.
\end{align}
It is worth to be noted that comparing Eq.\ \ref{eq:vgroup} and Eqs.\ \ref{eq:PTQW_eigenvectors} it is possible to verify that
\begin{equation}\label{eq:link_V_n}
   V_\delta(k) = n_z=-n_y
\end{equation}
This important result, which links some geometrical aspects of the QW system (that are used to define the system topology) and the dispersion relation, has been discussed widely in the main text. It is straightforward to show that there is a vector $\boldsymbol {a}$ perpendicular to $\boldsymbol{n}_{\delta}(k)$, 
for all values of $k$ and $\delta$; in other words, these eigenstates are positioned on a great circle of the Poincar\'e sphere, and such circle is independent of the value of $\delta$ (see Fig.\ \ref{fig:QW}). For a simpler comparison between the QW system and SSH model, we can choose $\boldsymbol {a}$ as the quantization axis for the coin ($z$ component). In the new representation, the single step operator $\op U_0'$ is obtained by means of a rotation of $\pi/4$ around the $x$-axis of the Poincar\'e sphere in the coin space. Accordingly the coin part of the associated eigenstates is confined on the equator of such sphere, that is $n'_z(k)=0$. It is straightforward to show that the effective Hamiltonian operator (\ref{hamiltonian}), with $\boldsymbol{n}_{\delta}(k)$ replaced by $\boldsymbol{n'}_{\delta}(k)$, has the same symmetries introduced for the SSH model, thus it belongs to the same topological class.\\

The SSH and QW systems we introduced have the same topological classification and may show distinct phases, according to the values of some external parameters. In SSH model, the quantum transition occurs when the hopping amplitude between different cells is equal to the one between the two sites of the same cell, that is $t=t^{\prime}$. In our quantum walk system, the phase transition occurs at $\delta=\pi/2 $ and $\delta=3\pi/2$, when $\sin{\frac{\delta}{2}}=\pm\cos{\frac{\delta}{2}}$, and the walker has the same probability either to remain in the same lattice site or to pass to the neighbour one.\\
It is straightforward to show 
that our QW model, when the localized Wannier wavefunction basis is adopted, 
can be mapped into a SSH-like model defined on a one-dimensional lattice where, in addition to the link 
between the two sites within the same cell, only hoppings between nonequivalent sites of different cells are present. 
However, unlike the SSH model, in the effective Hamiltonian not only nearest neighbor cells  connected: the hopping between nonadjacent cells is possible, that is the photon transfer integral associated with the orbital angular momentum displays a long-range behavior. This emerges when analyzing the property of the effective Hamiltonian, even though in such discrete-time systems the physically meaningful quantity to be considered is the single step evolution operator, which shows couplings only between nearest-neighbor cells.
\section{First and second order moments of the walker probability distribution: the large step-number limit for a QW system}\label{SI:Moments}
The large step-number limit for the walker distribution moments (Eqs.\ \ref{eq:PTQW_M1}-\ref{eq:PTQW_M2}) can be derived conveniently by evaluating $M_1$ and $M_2$ in momentum representation, where they are defined as follows:
\begin{align}
M_1&=\int_{-\pi}^\pi \frac{\text d k}{2\pi}\,\bra{\phi_0}\left(\hat U_0^\dag\right)^n (-i)\frac{\text d }{\text d k}\hat U_0^n\ket{\phi_0},\nonumber\\
M_2&=\int_{-\pi}^\pi \frac{\text d k}{2\pi}\,\bra{\phi_0}\left(\hat U_0^\dag\right)^n (-i)^2\frac{\text d^2 }{\text d k^2}\hat U_0^n\ket{\phi_0}.\label{eq:PTQW_dem}
\end{align}
Here $n$ is the step-number, $\hat U_0$ is the single step evolution operator (\ref{ev_operator}), and $\ket{\phi_0}$ is the coin initial state. Expanding the evolution operator as 
\begin{align}
\hat U_0^n&=\text{Exp}\left\{-i\,n\,E_\delta(k)\,\left[\boldsymbol{n}_{\delta}(k)\cdot\sigma\right]\right\} \nonumber \\
&=\cos\left\{n\,E_\delta(k)\right\}\,I_2-i\,\sin\left\{n\,E_\delta(k)\right\}\,\left[\boldsymbol{n}_{\delta}(k)\cdot\boldsymbol\sigma\right]
\label{eq:PTQW_dem}
\end{align}
where $I_2$ is the identity matrix in 2D, and the components of $\boldsymbol{n}_{\delta}(k)$ are those reported in Eq.\ \ref{eq:PTQW_eigenvectors}, it is straightforward to obtain the following equations:
\begin{align}
M_1/n&=\int_{-\pi}^\pi \frac{\text d k}{2\pi}\,V_\delta(k)\,\bra{\phi_0}\left(\boldsymbol{n}_{\delta}(k)\cdot\sigma\right)\ket{\phi_0}+ O(1/n),\label{eq:M1_SI}\\
M_2/n^2&=\int_{-\pi}^\pi \frac{\text d k}{2\pi}\,\left[V_\delta(k)\right]^2+ O(1/n^2),\label{eq:M2_SI}\\
\end{align}
Eq.\ \ref{eq:M1_SI}, reporting the expression for the first order moment, can be evaluated considering the $V_\delta=n_z=-n_y$ (Eq.\ \ref{eq:link_V_n}). Applying this substitution, the same equations reads
\begin{align}\label{eq:M1_SI2}
M_{1} /n&=  (s_{y}-s_{z}) \int_{-\pi}^\pi \frac{\text d k}{2\pi}\,\left[V_\delta(k)\right]^2+ O(1/n)
\end{align}
%
%
%
\begin{figure}[t]
\centering
\includegraphics[width=14.5cm]{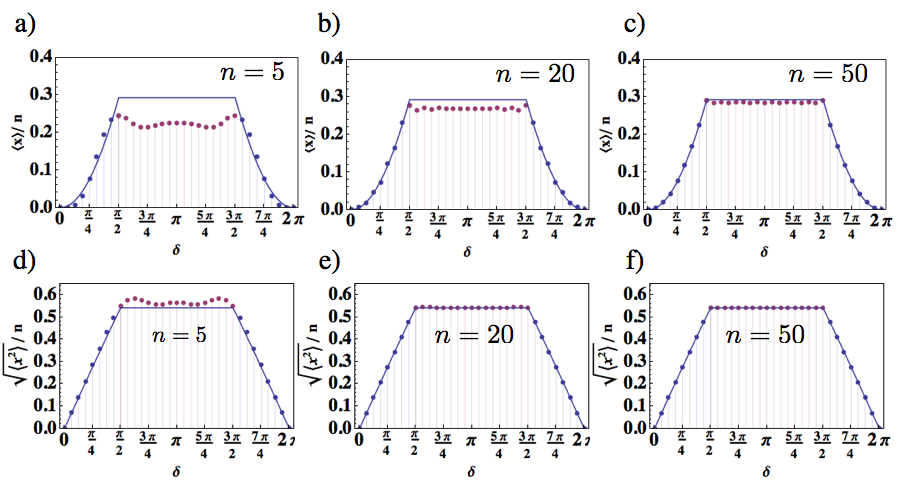}
\caption{Analysis of statistical moments $\{M_1,M_2\}$ in a QW. The system, initially prepared in the state $\ket{\psi_0}=\ket{\phi_0}_c\otimes\ket{0}_w$, undergoes a QW described by the step operator (\ref{ev_operator}). For every plot, purple points are obtained from a numerical simulation, when varying $\delta$ with steps of $\pi/16$ in the range $\{0,2\pi\}$; continuous blue lines represent the quantity $L(\delta)$ [panels a-c)], or $\sqrt{L(\delta)}$ [panels d-f)] (see Eq.\ \ref{eq:Ldeltafun}). For the simulation, we prepared the coin in the state $\{s_1,s_2,s_3\}=\{0,0,1\}$. Panels a-c) First order moment $M_1$, divided by the number of steps of the walk, as a function of the parameter $\delta$, for a walk of 5, 20, and 50 steps, respectively (this number is reported inside each figure). As $n$ increases, simulated data converge to the values predicted by Eq.\ \ref{eq:PTQW_M1}. Panels d-f) Square root of the second order moment, divided by the number of steps $n$. The figures are organized as in panels a-c). In this case, we can observe that simulated data converge much faster to the asymptotic values reported in Eq.\ \ref{eq:PTQW_M2}, with a discontinuity emerging even for a walk of few steps.} \label{fig:PTQW_M12}
\end{figure}
A plot for $L(\delta)$ (and for $\sqrt L$) is reported in Fig.\ \ref{fig:PTQW_M12}, where this asymptotic limit is compared to simulated data for a $n$-step QW, with $n={5,20,50}$. As discussed in the main text, it can be observed that as $n$ increases, both $M_1$ and $M_2$ converge to their asymptotical limit (\ref{eq:PTQW_M1},\ref{eq:PTQW_M2}), with this convergence being much faster for $M_2$ with respect to $M_1$.

First and second order moments reported in Eqs.\ \ref{eq:M1_SI2},\ref{eq:M2_SI} coincide with the expressions reported in the main text (Eqs.\ \ref{eq:PTQW_M1},\ref{eq:PTQW_M2}). In the specific case of our QW model, the integral appearing in both Eqs.\ \ref{eq:M1_SI2},\ref{eq:M2_SI} can be solved analytically.
Using the expression for the group velocity reported in Eq. \ref{eq:vgroup}, Eq.\ \ref{eq:PTQW_L} yields
\begin{equation}\label{eq:Ldelta}
  L(\delta) = \frac{1}{2\pi}\int_0^{2\pi}V_\delta^2dk = \frac{1}{2\pi}\int_0^{2\pi}\frac{\sin^2(\delta/2)\cos^2(k)}{2-[\cos(\delta/2)+\sin(\delta/2)\sin(k)]^2}dk.
\end{equation}
This integral can be calculated by the residue theorem passing to the complex variable $z = e^{ik}$. Then we obtain $L(\delta)=\oint f_\delta(z)dz$, where the integral is along the unit circle in the complex $z$-plane and $f_\delta(z)$ is given by
\begin{equation}\label{eq:f}
   f_\delta(z) = \frac{i (1+z^2)^2 \sin^2(\delta/2)}{\pi z [(1+z^2)^2 \cos(\delta)-z^4-10 z^2-1-4 i z (z^2-1) \sin(\delta)]}
\end{equation}
The poles of $f_\delta(z)$ are located on the imaginary axis at $z_k = \{0,i(\sqrt{2}-1)\cot(\delta/4),i(\sqrt{2}+1)\tan(\delta/4),-i(\sqrt{2}+1)\cot(\delta/4),-i(\sqrt{2}-1)\tan(\delta/4)\}$ $(k=1,\dots,5)$ and the residues of $f_\delta(z)$ at the poles are given by $2\pi i r_k=\{1,\frac{1}{4}[-\sqrt{2}+2\cos(\delta/2)],\frac{1}{4}[\sqrt{2}-2\cos(\delta/2)],\frac{1}{4}[\sqrt{2}+2\cos(\delta/2)],\frac{1}{4}[-\sqrt{2}-2\cos(\delta/2)]\}$, respectively. Apart from the pole at $z=0$, when $\delta$ varies, the locations of the poles move in the complex plane entering and exiting the unit circle, but only the residues of the poles inside the unit circle contribute to $L(\delta)$. Thus, we find
\begin{equation}\label{eq:Ldeltafun}
\arraycolsep=1.4pt\def\arraystretch{2.2}
  L(\delta) = \left\{\begin{array}{llll}
                    2\pi i(r_1+r_3+r_5)&=2\sin^2(\delta/4)  & \quad\mathrm{for}&0\le\delta\le\pi/2\\
                    2\pi i(r_1+r_2+r_5)&=1-\frac{1}{\sqrt{2}}& \quad\mathrm{for}&\pi/2\le\delta\le 3\pi/2\\
                    2\pi i(r_1+r_2+r_4)&=2\cos^2(\delta/4)  & \quad\mathrm{for}&3\pi/2\le\delta\le2\pi
              \end{array}\right.
\end{equation}

\section{Analysis of first and second order moments for the SSH model}\label{SI:SSH_moments}
The same approach we used for the analysis of statistical moments in the QW system can be used for the SSH model. In this case, we have that
\begin{align}\label{eq:SSH_M1}
\mathcal M_{1} /\tau&=  -s_{2} \mathcal L(t,t')+ O(1/\tau),
\end{align}
\begin{align}\label{eq:SSH_M2}
\mathcal M_{2}/\tau^2&= \mathcal L(t,t')+ O(1/\tau^2),
\end{align}
where we are considering $\tau$ as a continuous temporal coordinate. The quantity $\mathcal L$ has the same expression reported in Eq.\ \ref{eq:PTQW_L}, with the group velocity being that of the SSH model (Eq.\ \ref{eq:vgroup_ssh}). It can be noted that these expressions are obtained when considering an electron prepared in a localized initial state. While this is a standard choice for QWs, in electron dynamics this condition is hard to be reproduced experimentally, and typically is not considered. Even in this case, $\mathcal L$ has an analytical expression. Passing to the complex variable $z=\text e^{i\,k}$, we find
\begin{align}\label{eq:SSH_res}
\mathcal L=-\frac{i\,tt'}{8\pi}\oint \text dz\,\frac{(z^2-1)^2}{z^2\left[z^2-z\left(\frac{t'^2+t^2}{tt'}\right)+1\right]}
\end{align}
where the integral is evaluated along the unit circle $|z|=1$ in the complex plane. Eq.\ \ref{eq:SSH_res} can be solved by residue method. The integrand function has three poles, which are located along the real axis:
\begin{align}\label{eq:SSH_poles}
z_0=0;\qquad z_1=t/t'; \qquad z_2=t'/t;
\end{align}
\begin{figure}[t]
\begin{center} 
\includegraphics[width=9cm]{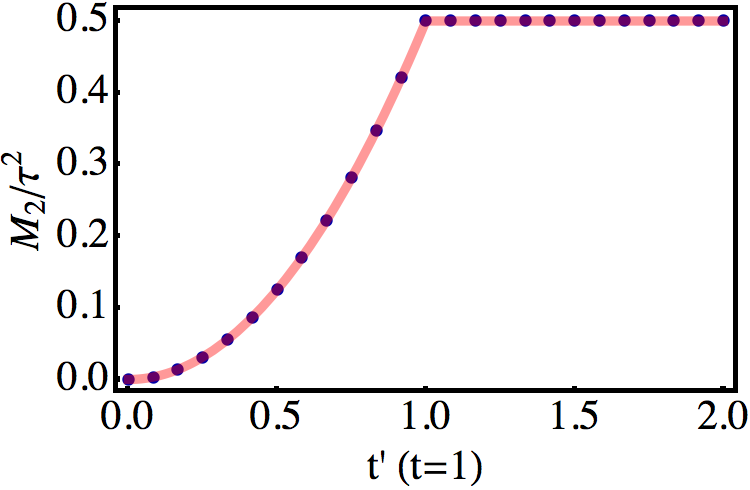} 
\caption{Second order moment for the SSH model in the large time limit. We plot the second order moments $\mathcal M_2$ corresponding to the probability distribution for a single electron whose dynamics is ruled by the SSH Hamiltonian (\ref{eq:hamiltonian_ssh}). As for the QW, we considered a localized initial state (at $\tau =0$). Blue circles represent the data obtained from a numerical simulation, in which we evolved the initial wave function at the time $\tau=50$ (we recall that we are dealing with adimensional units). In such simulation, we considered $t$ as a constant parameter (i.e. $t=1$), while varying $t'$ within the interval $\{0,2\}$ with steps of $2/25$. The red continuous line represents the asymptotic limit calculated using Eqs.\ \ref{eq:SSH_M2} and \ref{eq:Ldeltafun_SSH}. We can see that such limit reproduces well the simulated data.}
\label{fig:SSH_res}
\end{center}
\end{figure}
It is worth noticing that $z_0$ is doubly degenerate. When varying $t$ and $t'$, poles $z_1$ and $z_2$ move in the complex plane, entering or exiting the unit circle. Since only inner poles contribute to the integral (\ref{eq:SSH_res}), residues at $z_1$ and $z_2$ should not be considered simultaneously; thus, in the expression of $\mathcal L$ the residue at $z_1$ $(z_2)$ will appear when $t<t'$ ($t>t'$). If $t=t'$, $z_1$ and $z_2$ are located on the integration path (the unit circle) and the Eq.\ \ref{eq:SSH_res} cannot be solved using this method. The residues at the poles $z:k$ reported in Eq.\ \ref{eq:SSH_poles} are 
\begin{align}\label{eq:SSH_residues}
2\pi i\, r_0=\frac{t^2+t'^2}{4};\qquad 2\pi i\, r_1=\frac{t^2-t'^2}{4}; \qquad 2\pi i\, r_2=\frac{t'^2-t^2}{4}.
\end{align}
Accordingly, we have that
\begin{align}\label{eq:Ldeltafun_SSH}
\arraycolsep=1.4pt\def\arraystretch{2.2}
  \mathcal L(t,t') = \left\{\begin{array}{ccc}
                    2\pi i(r_0+r_2)=\frac{t'^2}{2}  &\quad \mathrm{for}&t'<t\\
		   2\pi i(r_0+r_1)=\frac{t^2}{2}  &\quad \mathrm{for}&t'>t\\              \end{array}\right.
\end{align}
In Fig.\ \ref{fig:SSH_res} we plot the function $\mathcal L$, and we compare it to simulated results for the evolution of the electron dynamics described by the SSH model. In the large time limit, numerical data converge rapidly to $\mathcal L$ (\ref{eq:Ldeltafun_SSH}). 
It is important to observe that, as a difference with respect to the QW model, here $M_2$ would vary in the non-trivial phase too, for any variation of the parameter $t$, that we considered as a constant. The observable quantity that is locked to a constant in the non-trivial phase, independently of the hopping parametrization, is $M_2/t^2$. Remarkably, as for the QW system, this is equal to $n_y^2$, integrated over the Brillouin zone in momentum space (see Eq.\ \ref{eq:n_V_SSH}). 
\begin{thebibliography}{10}

\bibitem{Xiao10_RevModPhys}
D.~Xiao, M.~Chang, and Q.~Niu.
\newblock Berry phase effects on electronic properties.
\newblock {\em Rev. Mod. Phys.}, 82:1959--2007, Jul 2010.

\bibitem{Qi11_RevModPhys}
X.~Qi and S.~Zhang.
\newblock Topological insulators and superconductors.
\newblock {\em Rev. Mod. Phys.}, 83:1057--1110, Oct 2011.

\bibitem{Rudn09_PRL}
M.~Rudner and L.~Levitov.
\newblock Topological transition in a non-hermitian quantum walk.
\newblock {\em Phys. Rev. Lett.}, 102:065703, Feb 2009.

\bibitem{Kita10_PRA}
T.~Kitagawa, M.~S. Rudner, E.~Berg, and E.~Demler.
\newblock Exploring topological phases with quantum walks.
\newblock {\em Phys. Rev. A}, 82:033429, 2010.

\bibitem{Kita12_NatCom}
T.~Kitagawa, M.~A. Broome, A.~Fedrizzi, M.~S. Rudner, E.~Berg, I.~Kassal,
  A.~Aspuru-Guzik, E.~Demler, and A.~G. White.
\newblock Observation of topologically protected bound states in photonic
  quantum walks.
\newblock {\em Nat. Commun.}, 3(882), 2012.

\bibitem{Asbo12_PRB}
J.~K. Asb\'oth.
\newblock Symmetries, topological phases, and bound states in the
  one-dimensional quantum walk.
\newblock {\em Phys. Rev. B}, 86:195414, Nov 2012.

\bibitem{SSH_PRL}
W.~Su, J.~Schrieffer, and A.~Heeger.
\newblock Solitons in polyacetylene.
\newblock {\em Phys. Rev. Lett.}, 42:1698--1701, Jun 1979.

\bibitem{Thou82_PRL}
D.~J. Thouless, M.~Kohmoto, M.~P. Nightingale, and M.~den Nijs.
\newblock Quantized hall conductance in a two-dimensional periodic potential.
\newblock {\em Phys. Rev. Lett.}, 49:405--408, Aug 1982.

\bibitem{Zhan05_Nat}
Yuanbo Zhang, Yan-Wen Tan, Horst~L. Stormer, and Philip Kim.
\newblock Experimental observation of the quantum hall effect and berry's phase
  in graphene.
\newblock {\em Nature}, 438(7065):201--204, 11 2005.

\bibitem{Hasa10_RevModPhys}
M.~Z. Hasan and C.~L. Kane.
\newblock \textit{Colloquium} : Topological insulators.
\newblock {\em Rev. Mod. Phys.}, 82:3045--3067, Nov 2010.

\bibitem{Gome12_Nat}
K.~K. Gomes, W.~Mar, W.~Ko, F.~Guinea, and H.~C. Manoharan.
\newblock Designer dirac fermions and topological phases in molecular graphene.
\newblock {\em Nature}, 483(7389):306--310, 03 2012.

\bibitem{Atal13_NatPhys}
M.~Atala, M.~Aidelsburger, Barreiro~J. T., D.~Abanin, T.~Kitagawa, E.~Demler,
  and I~Bloch.
\newblock Direct measurement of the zak phase in topological bloch bands.
\newblock {\em Nature Phys.}, 9:795--800, 2013.

\bibitem{Gens13_PRL}
M.~Genske, W.~Alt, A.~Steffen, A.~Werner, R.~Werner, D.~Meschede, and
  A.~Alberti.
\newblock Electric quantum walks with individual atoms.
\newblock {\em Phys. Rev. Lett.}, 110:190601, May 2013.

\bibitem{Hauk12_PRL}
P.~Hauke, O.~Tieleman, A.~Celi, C.~\"Olschl\"ager, J.~Simonet, J.~Struck,
  M.~Weinberg, P.~Windpassinger, K.~Sengstock, M.~Lewenstein, and A.~Eckardt.
\newblock Non-abelian gauge fields and topological insulators in shaken optical
  lattices.
\newblock {\em Phys. Rev. Lett.}, 109:145301, Oct 2012.

\bibitem{Zeun14_Axv}
Julia~M. Zeuner, Mikael~C. Rechtsman, Yonatan Plotnik, Yaakov Lumer, Mark~S.
  Rudner, Mordechai Segev, and Alexander Szameit.
\newblock Probing topological invariants in the bulk of a non-hermitian optical
  system.
\newblock {\em arxiv:1408.2191 [quant-ph]}, 08 2014.

\bibitem{Asbo13_PRB}
J.~K. Asb\'oth and H.~Obuse.
\newblock Bulk-boundary correspondence for chiral symmetric quantum walks.
\newblock {\em Phys. Rev. B}, 88:121406, Sep 2013.

\bibitem{Vene12_QIP}
S.~E. Venegas-Andraca.
\newblock Quantum walks: a comprehensive review.
\newblock {\em Quantum Information Processing}, 11(5):1015--1106, 2012.

\bibitem{Wang13}
J.~Wang and K.~Manouchehri.
\newblock {\em Physical Implementation of Quantum Walks}.
\newblock Springer, 2013.

\bibitem{Card15_SciAdv}
Filippo Cardano, Francesco Massa, Hammam Qassim, Ebrahim Karimi, Sergei
  Slussarenko, Domenico Paparo, Corrado de~Lisio, Fabio Sciarrino, Enrico
  Santamato, Robert~W. Boyd, and Lorenzo~and Marrucci.
\newblock Quantum walks and wavepacket dynamics on a lattice with twisted
  photons.
\newblock {\em Science Advances}, 1(2), 03 2015.

\bibitem{Yao11_AOP}
Alison~M. Yao and Miles~J. Padgett.
\newblock Orbital angular momentum: origins, behavior and applications.
\newblock {\em Adv. Opt. Photon.}, 3(2):161--204, Jun 2011.

\bibitem{Marr06_PRL}
L.~Marrucci, C.~Manzo, and D.~Paparo.
\newblock Optical spin-to-orbital angular momentum conversion in inhomogeneous
  anisotropic media.
\newblock {\em Phys. Rev .Lett.}, 97:163905, 2006.

\bibitem{Picc10_APL}
B.~Piccirillo, V.~D'Ambrosio, S.~Slussarenko, L.~Marrucci, and E.~Santamato.
\newblock Photon spin-to-orbital angular momentum conversion via an
  electrically tunable q-plate.
\newblock {\em Appl. Phys. Lett.}, 97:241104, 2010.

\bibitem{Ryu10_NJP}
Shinsei Ryu, Andreas~P Schnyder, Akira Furusaki, and Andreas~WW Ludwig.
\newblock Topological insulators and superconductors: tenfold way and
  dimensional hierarchy.
\newblock {\em New Journal of Physics}, 12(6):065010, 2010.

\bibitem{Schn08_PRB}
Andreas~P. Schnyder, Shinsei Ryu, Akira Furusaki, and Andreas W.~W. Ludwig.
\newblock Classification of topological insulators and superconductors in three
  spatial dimensions.
\newblock {\em Phys. Rev. B}, 78:195125, Nov 2008.

\bibitem{Kita09_AIP}
Alexei Kitaev.
\newblock Periodic table for topological insulators and superconductors.
\newblock {\em AIP Conf. Proc.}, 1134(1), 2009.

\bibitem{Altl97_PRB}
Alexander Altland and Martin~R. Zirnbauer.
\newblock Nonstandard symmetry classes in mesoscopic normal-superconducting
  hybrid structures.
\newblock {\em Phys. Rev. B}, 55:1142--1161, Jan 1997.

\end{thebibliography}
\end{document}